\documentclass{article}

\usepackage{arxiv}
\usepackage{amsmath,amsfonts}
\usepackage{algorithmicx}
\usepackage{algorithm}
\usepackage{array}
\usepackage[caption=false,font=normalsize,labelfont=sf,textfont=sf]{subfig}
\usepackage[utf8]{inputenc} 
\usepackage[T1]{fontenc}    
\usepackage{hyperref}       
\usepackage{url}            
\usepackage{booktabs}       
\usepackage{amsfonts}       
\usepackage{nicefrac}       
\usepackage{microtype}      
\usepackage{lipsum}
\usepackage{graphicx}
\graphicspath{ {./images/} }

\usepackage[authoryear,longnamesfirst]{natbib}
\usepackage{textcomp}
\usepackage{stfloats}
\usepackage{verbatim}
\usepackage{cite}
\usepackage{enumerate}
\usepackage{mathtools}
\usepackage{algpseudocode}
\usepackage{float}

\newcommand\blfootnote[1]{%
  \begingroup
  \renewcommand\thefootnote{}\footnote{#1}%
  \addtocounter{footnote}{-1}%
  \endgroup
}

\title{A Spreader Ranking Algorithm for Extremely Low-Budget Influence Maximization in Social Networks using Community Bridge Nodes}

\author{
 Aaryan Gupta\footnotemark[1] \\
  Biometric Research Laboratory\\
  Delhi Technological University\\
  New Delhi, India\\
  \texttt{aryan227227@gmail.com} \\
   \And
 Inder Khatri\footnotemark[1] \\
  Biometric Research Laboratory\\
  Delhi Technological University\\
  New Delhi, India\\
  \texttt{inderkhatri999@gmail.com} \\
  \And
 Arjun Choudhry\footnotemark[1] \\
  Biometric Research Laboratory\\
  Delhi Technological University\\
  New Delhi, India\\
  \texttt{choudhry.arjun@gmail.com} \\
  \And
  Pranav Chandhok \\
  Delhi Technological University\\
  New Delhi, India\\
  \texttt{pranav.chandhok2000@gmail.com} \\
  \And
  Dinesh Kumar Vishwakarma \\
  Biometric Research Laboratory\\
  Delhi Technological University\\
  New Delhi, India\\
  \texttt{dinesh@dtu.ac.in} \\
  \And
  Mukesh Prasad \\
  School of Computer Science \\
  University of Technology Sydney\\
  Ultimo, Australia\\
  \texttt{mukesh.prasad@uts.edu.au} \\ 
}

\begin{document}
\maketitle
\begin{abstract}
In recent years, social networking platforms have gained significant popularity among the masses like connecting with people and propagating one's thoughts and opinions. This has opened the door to user-specific advertisements and recommendations on these platforms, bringing along a significant focus on Influence Maximisation (IM) on social networks due to its wide applicability in target advertising, viral marketing, and personalized recommendations. The aim of IM is to identify certain nodes in the network which can help maximize the spread of certain information through a diffusion cascade. While several works have been proposed for IM, most were inefficient in exploiting community structures to their full extent. In this work, we propose a community structures-based approach, which employs a K-Shell algorithm in order to generate a score for the connections between seed nodes and communities for low-budget scenarios. Further, our approach employs entropy within communities to ensure the proper spread of information within the communities. We choose the Independent Cascade (IC) model to simulate information spread and evaluate it on four evaluation metrics. We validate our proposed approach on eight publicly available networks and find that it significantly outperforms the baseline approaches on these metrics, while still being relatively efficient. \blfootnote{*Equal Contribution}
\end{abstract}

\section{Introduction}\label{sec1}

Complex Network Analysis can be understood as the analysis of a primary structure of a network and its dynamical characteristics \citep{CN1,CN2}. In these frameworks, the elements that build these networks are addressed as nodes and the connections or relations between them are addressed as edges. For example, in a reference framework, publications form the nodes, while their citations form the edges. In an internet-based social networking platform, a person's records represent a node, while the person's various connections with other people are represented by other nodes (like \emph{kinship} or \emph{follow-supporter}) form the edges. Generally, these genuine organizations observe \emph{Power-Law Degree Circulation}, which makes their conduct eccentric and their study back-breaking \citep{scalefree}.

Online Social Networks (OSNs) provide a convenient way for users to collaborate and convey their ideas to other people \citep{OSN1,OSN2,OSN3}. With the web insurgency, the unexpected upsurge in the number of users on OSNs has given the showcasing experts yet another methodology to consider, which can be more compelling and less expensive than the conventional methodologies. This methodology, also known as \emph{Viral Marketing} \citep{viralmarketingFerguson}, employs OSNs to acquire a significantly larger audience, as compared to the traditional methods. One way of performing viral marketing is Influence Maximisation, which identifies a given number of influential nodes in a network, which induces the maximum spread possible.

In recent years, several approaches have been proposed for Influence Maximisation due to the task's significant importance and applicability in the industrial domain. A lot of these approaches are local and semi-local structure-based algorithms that rely on the different local aspects of the network. Apart from these, various other works have relied on the global properties of the network, which generally require a longer time for execution. Further, some works consider \emph{bridge nodes} to be the most influential nodes and pose Influence Maximisation (IM) as a bridge node identification problem. Some recent works also try to identify the core nodes as influential nodes. 

Recently, various community structure-based approaches have been proposed, which divide the network into small sub-groups called communities, and further, find spreaders by using these structures. These community-based algorithms are found to be more efficient and simpler for real-life utilization. However, these approaches focus on extracting the core nodes, which have a good influence over their respective communities. However, in the case of extremely low-budget IM, where the number of spreaders is extremely small, it is likely that few communities get influenced using these approaches. To counter this, we instead propose the use of community bridge nodes as spreader nodes, which have connections to a large number of communities and can aid in maximising the influence spread in the case of a very small number of initial spreaders.

In this paper, we present a novel method for tackling Influence Maximisation in complex networks using a novel centrality measure that uses the concepts of community structures and K-shell Decomposition to identify the influential spreaders in a network. Our method scores a node by considering its connection to different communities, and further scores the strength of its relation to those connected communities. We define three measures, i.e., Community K-Shells, Community K-Shell Entropy, and Community K-Shell Score (CKS\_Score), which are used to evaluate the connection of a node to the communities in terms of their qualitative and quantitative aspects, and finally, rank the nodes on the basis of these measures. The measures are able to identify and give a higher score to the nodes which have the maximum spreading capacity or influence over the entire network. For further experimental analysis to verify our proposed approach, we use the Independent Cascade (IC) model to simulate the influence propagation on eight different networks. Our experimental evaluations over various datasets and performance metrics validate the superiority of our proposed approach over seven previous state-of-the-art approaches across different domains and act as evidence of its efficacy and widespread applicability. The main contribution of this paper can be summarised as:
\begin{itemize}
    \item We propose a novel approach for Influence Maximisation in low-budget scenarios by employing the concepts of Community Structure, K-Shell Decomposition, and Shannon's Entropy. 
    \item We define three novel measures, namely Community K-Shells, Community K-Shell Entropy, and CKS\_Score, that help to qualitatively and quantitatively evaluate the connections of a node to various communities.
    \item We verify our approach by evaluating it on eight different datasets using various performance metrics, and further compare it with the existing approaches. We also perform the Friedman and Iman-Davenport statistical tests to verify the superiority of our approach.
\end{itemize}

The remainder of the paper is presented as follows. Section 2 consists of the related works and the existing research gaps. Section 3 contains preliminary information needed to evaluate and understand our approach. Section 4 contains our proposed methodology and algorithms. Section 5 contains information about the various experimental details like the datasets used, baseline models, the performance metrics used, and the implementation environment. Section 6 consists of our experimental results, their analysis, and broad outcomes. Section 7 contains the Friedman and Iman-Davenport statistical tests to verify the superiority of our approach. Section 8 contains our concluding statement.

\section{Related Works \& Research Gaps}\label{sec2}

Over the last few decades, Influence Maximisation as a task has seen a significant number of contributions from researchers. A wide variety of methods have been proposed for finding the most viable and influential spreader nodes \citep{IMsurvey,IMsurvey2,IMsurvey3}. Among these approaches, centrality measure-based methods have recently garnered significant attention from researchers due to their reduced time complexity and simpler methodology \citep{IMCentralitysurvey,IMCentralitysurvey2}. These approaches usually consist of two steps: generating a score for each node using a defined centrality measure and selecting the top \textit{k} on the basis of the calculated score. 

\citet{Freeman1978CentralityIS} proposed one of the earliest methods involving Degree Centrality, which directly considers the number of neighbours to generate a score for a node. Degree Centrality has linear time complexity and was found to be effective in small networks. Due to its dependence only on the immediate neighbours of the node for scoring, the efficacy of Degree Centrality is questionable while scoring the nodes with lesser but highly influential neighbour nodes.

As an alternative to degree centrality, some semi-local centrality measures like Local Centrality (LC) \citep{LC} and Local Structural Centrality (LSC) \citep{LSC} have also been introduced, which go beyond the immediate neighbours but up to a limited range. LC ranks these on the basis of one-hop and two-hop nearest neighbours, whereas LSC further considers an additional factor, i.e. the clustering coefficient between the immediate neighbours. Some semi-local methods also try to identify the bridge nodes in a network as influential nodes. Bridge nodes are the nodes which connect different groups of nodes. \citet{DIL} defined the Degree and Importance of Lines (DIL) centrality measure, which uses the degree value and the importance of lines to highlight the bridge node. It further takes into account the number of connected triangles when determining the relevance of a link, with the bridge node taking the central role in controlling information propagation. \citet{DCL} proposed Dense and Centrality Localization (DCL), a measure which uses a node's degree, the clustering coefficient, and the relationship between its one-hop neighbours to find out the bridge nodes. \citet{LID} proposed Local Information Dimensionality (LID), an approach which identifies influencers in complex networks by considering quasi-local information, i.e., local structural properties. However, since these approaches do not take into account a node's global location and only use the information corresponding to a limited area or locality, their implications in real-life large networks become less efficient.

Some recent works have also used global centrality measures like Betweenness Centrality, Closeness Centrality, and K-shell Decomposition, among others. \citet{BC} defined Betweenness for a node as the ratio of the shortest path which passes through the given node, and the total number of possible shortest paths in the network. \citet{CC} proposed Closeness for a node as the inverse of the sum of the shortest path to all other nodes, hereby determining the average closeness of a node to the rest of the graph. As the calculation of Betweenness and Closeness requires the calculation of the shortest path between every node pair of the network, the time complexity becomes significantly high, which is undesirable. Eigenvector Centrality \citep{EVC} and Pagerank Centrality \citep{PR} measures are some other Global centrality measures used for IM. These measures score the nodes on the basis of their importance based on their neighbourhood nodes. According to these measures, the presence of important nodes in the surroundings makes the centre node important.

\citet{kshell} proposed K-shell decomposition, another global centrality measure, which divides the nodes into different levels on the basis of their location with respect to the core of the network. Normally, the nodes with the highest shell number are nearest to the core and are selected as the initial seeds. As K-shell decomposition assigns more than one node to a given shell number, this often makes it unreliable as the nodes still in the same shell can produce different influences.

\citet{ENC} proposed an improved algorithm called the Extended Neighbourhood Coreness (ENC) measure, which ranks nodes on the basis of the sum of the K-shell of the neighbourhood nodes. This concept was further extended to get Extended Neighbourhood Coreness by taking the sum of the Neighbourhood Coreness values of a node's neighbours. \citet{MDD} introduced Mixed Degree Decomposition (MDD), an iteration of K-shell that further distinguishes nodes in the same shell by using the removed and residual neighbours during the K-shell decomposition process. 

In recent years, a lot of works have been proposed that take the help of the community structures to find out the spreader nodes that could maximise the influence spread in the network. Some of the most relevant community-based works from the last few years are Community-based Approach for Opinion Maximization (CAOM) \citep{CAOM}, Community-based algorithm for Finding Influential Nodes (CFIN) \citep{CFIN}, and Gateway Local Rank (GLR) \citep{GLR}. The community-based works focus on finding out the core nodes in the significant communities, and further leverage the higher connection density of communities to increase the spread using these nodes as initial spreaders. CAOM divides the network into community structures, and the spreaders are divided among the most significant communities. The candidate nodes are hereafter generated in significant communities using a one-hop measure and further nodes are shortlisted by using a two-hop measure and \emph{Elimination of Overlapping Influence} policy. Like CAOM, CFIN also finds the candidate nodes among the significant communities after community detection. The initial nodes are selected in significant communities with the help of degree centrality, and the seed nodes are finally selected by applying local clustering coefficients over the initial nodes. GLR finds out the prominent nodes in the communities by simplifying the closeness centrality for the community. The best core nodes, i.e., the local critical node and the gateway node are found in each community. Then, the shortest paths of all other nodes to the best core nodes are found and the nodes are given scores based on their closeness to core nodes.

We find that most of the previous works based on the communities consider the core nodes from the significant communities to be the influential spreaders. These core nodes usually have good control over the respective communities in which they are present, and can influence a large fraction of community nodes when used as initial spreaders. In the case of real-life applications of influence maximization, where the number of spreaders is very small, it is likely that only a few selected communities would be influenced, which can be a serious drawback in the long run. To overcome such scenarios, we try to find out the community bridge nodes which are connected to a large number of communities and can simultaneously propagate information to these connected communities. There are a few other recent works like LID, DCL, and DIL, which also use bridge nodes as initial spreaders. However, these methods rely upon local information, and the selected node might not have influence beyond a given locality. Contrary to these approaches, our proposed approach relies upon global metrics like K-shell Decomposition and community structure to find out the community bridge nodes, which can act like global bridge nodes.

\section{Preliminaries}\label{prelim}

This section contains preliminary information which will be utilized in our approach. We share details about the influence maximization task, the uses of community structures in social network tasks, K-shell decomposition, and the Information Propagation model used.

\subsection{Influence Maximisation}
Influence Maximisation (IM) can be defined as the task of distinguishing \textit{k} potential nodes whose impact spread over the entire organization is most effective for any association of \textit{N} users \citep{kempeIM}. The aim is to boost the spread of information diffusion and increase the number of nodes affected. The initial step in IM is to pick \emph{k} most influential spreader nodes. However, due to a significant constraint on assets and time, the number of seed nodes should be fundamentally restricted in comparison to the entire populace. Formally, IM can be considered as: $G(V,E)$, where $V$ is the set of nodes, $E$ is the set of edges, and $k$ is the certain number of seed nodes. IM refers to selecting a set of $k$ nodes such that they maximize the influence spread. The influence spread $S'$ from a set of seed nodes $S$ is maximum at the end of the diffusion process and is denoted as $\sigma (S)$. Mathematically, it can be presented as shown below.
\begin{equation}\label{eq:im}
S'= \arg\max \sigma(S)\ where\
S \subset V, |S| = k
\end{equation}

\subsection{Community Structure}
A \emph{community} is a part of a network that comprises nodes with similar behavior and characteristics. For example, if nodes are symbolized as people, then people in the same age group, or friends circle, or with the same home zip codes are treated as a community. The connections within these nodes are stronger as compared to the connections between these nodes and the rest of the network. Usually, connections within a community are dense, whereas the connections between nodes of different communities are relatively sparse. Therefore, it is assumed that a piece of information propagates efficiently and quickly when injected in a community \citep{Comm1, ES_Comm, ES_Comm2, ES_Comm3}. Community structures are useful entities that aid not just in Influence Maximisation \citep{ES_Comm2}, but also in Link Prediction \citep{ES_LP}, Viral Marketing \citep{viralmarketingFerguson}, Epidemic Prevention Management \citep{ES_EPM} and other tasks.

\subsection{K-Shell Decomposition}
K-Shell Decomposition \citep{kshell} can be formally defined as the global centrality measure to divide nodes in accordance with their degree and their vicinity to the core of the network. 

It comprises the following steps:
\begin{itemize}
    \item Every node in the network is supposed to be assigned to a K-shell number.
    \item The nodes with degree one are recurrently removed from the network till only the higher-degree nodes are left. The removed nodes are allotted to the degree one shell.
    \item The process is repeated for higher degrees i.e. two, three, and so on, with the allotment of the respective shells until the highest degree, i.e., the degree of the core, is achieved. 
\end{itemize}

The nodes with higher K-Shell values are closer to the core of the network and have a higher degree. These nodes are also considered more influential in comparison to the nodes with lower shell values.

\subsection{Information Propagation Model}
Information propagation (IP) is the process of spreading a piece of information in a complex network \citep{IP1,IP2}. It is generally used to evaluate the different spreader-ranking methods. To model information propagation, epidemic models are used. Here, the nodes are treated as a population, and information is modeled as an infectious disease that spreads in the population on interaction. Various epidemic models have been used in recent years for IP processes including Susceptible-Infected-Recovered (SIR) \citep{SIR} model, Independent Cascade (IC) \citep{IC}, and Linear Threshold (LT) \citep{LT}. In this manuscript, we use the Independent Cascade model (IC) to evaluate our proposed approach CKS and further compare it with other state-of-the-art techniques. In the IC model, the nodes can be assigned either one of the two states, i.e., infected nodes or susceptible nodes. Information propagation starts by initially assigning a few seed nodes as infected nodes, which further interact and try to infect other neighboring nodes. Similarly, every infected node tries to infect its neighbor and the process continues till there are no new nodes infected. There is some probability associated with each interaction which describes the chances of infecting a susceptible node. Finally, the total influence of a seed node is estimated by taking the sum of all the nodes which are infected by the end of information diffusion.

\begin{figure}[b!]
\centering
\includegraphics[scale=0.75]{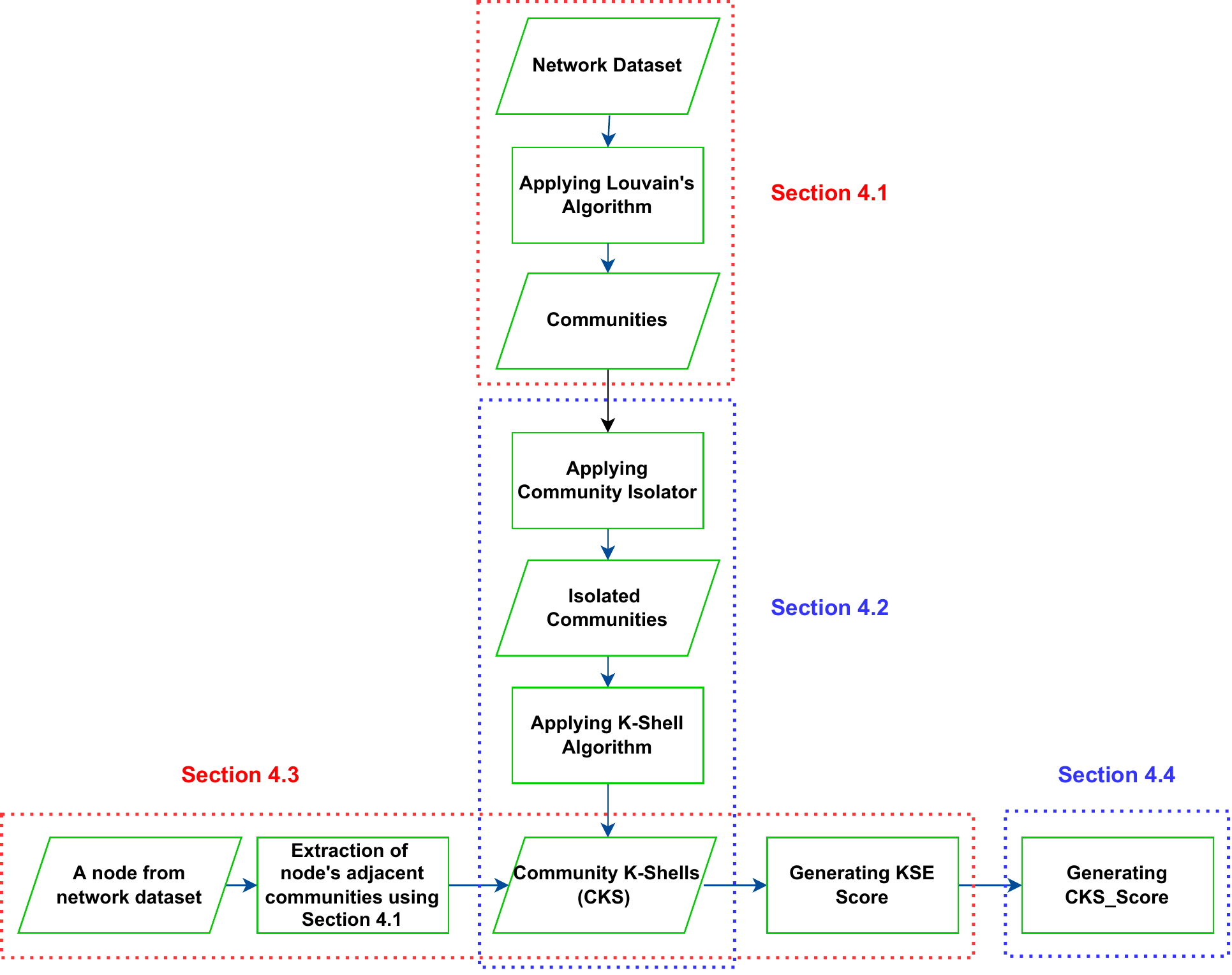}
\caption{Flowchart representation of the steps involved for the calculation of CKS\_Score.} 
\label{fig:cks_score-steps}
\end{figure}

\section{Proposed Methodology}\label{proposed}
In this section, we propose Community K-Shells Score (CKS\_Score), a novel algorithm for finding out the seed nodes for Influence Maximization in complex networks. This approach utilizes the concept of communities, K-shell decomposition, and Shannon's Entropy to find out the bridge nodes which are capable of simultaneously spreading information to different subgroups of the network. 

Contrary to most of the previous approaches which rank nodes on the basis of their connection to other nodes, we evaluate the nodes on the basis of their connection to the different communities present in the network. We further ensure that the information is being spread in a well-structured manner throughout the community, not just a small part of it. We first compute the KSE score for the connections between a node and the corresponding communities using pre-calculated Community K-Shells. This score depicts how well a piece of information would be transmitted to the community if the given node is informed about it. Then, we compute the CKS\_Score for a node by aggregating all the KSE scores between the respective node and connected communities. These aggregated scores are simultaneously weighed by the number of connections between the node and the community, and also the size of the community. 

The procedure for the calculation of CKS\_Score, represented pictorially in Fig. \ref{fig:cks_score-steps}, is as follows:

\begin{enumerate}
\item Detecting the communities in the network using the Louvain algorithm \citep{louvain}, which is a partial multi-level algorithm to detect communities in large datasets.
\item Obtaining the Community K-Shells (CKSs) by using the K-shell centrality measure over the results obtained from the Louvain algorithm.
\item Obtaining the Community K-Shell entropy using the CKSs obtained, for the connections between a node and its connected community, which results in nodes with strong distributed connections to a community.
\item Computing the CKS\_Score to obtain nodes with strong distributed connections to multiple significant communities.
\end{enumerate}

\subsection{Identifying communities}
Our proposed methodology begins by identifying communities in the network dataset, which gives a better way of tracking the flow of information in the network. 

In this work, we use the Louvain algorithm \citep{louvain} to detect the communities. Louvain’s algorithm is based on greedy optimization techniques, which results in efficient processing and quicker convergence.  It works in a step-by-step manner of repeating 2 phases i.e. Local moving of nodes and Aggregation of the network.  We use this algorithm for the following reasons: 

\begin{enumerate}
    \item It considers the size of communities, hence it ends up selecting only the impactful communities.
    \item It tries to maximize the difference between the actual number of edges in a community, and the expected number of edges in the community.
    \item It is based on greedy optimization techniques, which helps in faster convergence and hence makes the process very efficient.
\end{enumerate}

\subsection{Obtaining Community K-Shells}
\label{section-cks}
Finding Community K-Shells (CKS) is a crucial step in determining the influence of a node in its respective community. The algorithm for obtaining CKS consists of two major steps: isolating the communities and applying the K-Shell algorithm over the isolated community graph.

Isolating the communities helps the K-Shell algorithm handle each community separately. Since the K-Shell algorithm considers the internal structure of the network, applying it over individual communities aids in the analysis of deeper internal structures of the communities. Therefore, it is able to analyze networks not just at a macro level, but also at a micro level by tackling and handling communities individually.

 \begin{itemize}
    \item The isolation of the communities involves deleting the edges between nodes from different communities. Hence, we achieve a graph with several unconnected sub-graphs representing various communities. Any computation over these community sub-graphs will be handled by the algorithm as if each of these communities represents a separate graph.
    \item While applying the K-Shell algorithm to these sub-graphs, the isolated communities are passed through the K-Shell algorithm, which divides the nodes of a community into various shells as per their closeness to the core of the community. The closer a shell is to the core, the greater the influence its constituent nodes will have on the respective community. Each shell is assigned a specific community k-value, which is equal to 1 for the outermost shell, and it increases as we move toward the core of the community. Therefore, the greater the k-value of a shell, the higher the impact of its constituent nodes on the respective community.
\end{itemize}

\begin{algorithm}[h!]
 \caption{: Community K-Shell}
 \begin{algorithmic}[1]
 \label{algo:1}
 \renewcommand{\algorithmicrequire}{\textbf{Input:}}
 \renewcommand{\algorithmicensure}{\textbf{Output:}}
 \Require $G = (V,E)$
  \Ensure  $Community\_K-Shell$
    \State $\{c_1,c_2,c_3....c_m\} \leftarrow G\ partitioned\ using\ the \ Louvain\ $ $ Algorithm$
    \State $H \leftarrow Isolated\ communities\ by\ deleting\ inter-$ $community\ connections\ in\ G$
    \State $CKS \leftarrow Apply\ K-shell\ algorithm\ over\ H$\\
 \Return $CKS$
 \end{algorithmic} 
\end{algorithm}

Algorithm 1 represents the algorithmic procedure used to compute Community K-Shell (CKS) measure for graph G.
\begin{itemize}
\item We divide the graph into various communities using the Louvain algorithm. 
\item Then, the edges connecting the nodes from different communities are removed, which returns a graph consisting of isolated communities. 
\item Then, we execute the K-shell algorithm over the already-segregated community graph, which returns the Community K-Shell for each node $v$.\\
\end{itemize}

\begin{figure*}[h!]
\centering
\centerline{\includegraphics[scale=0.8]{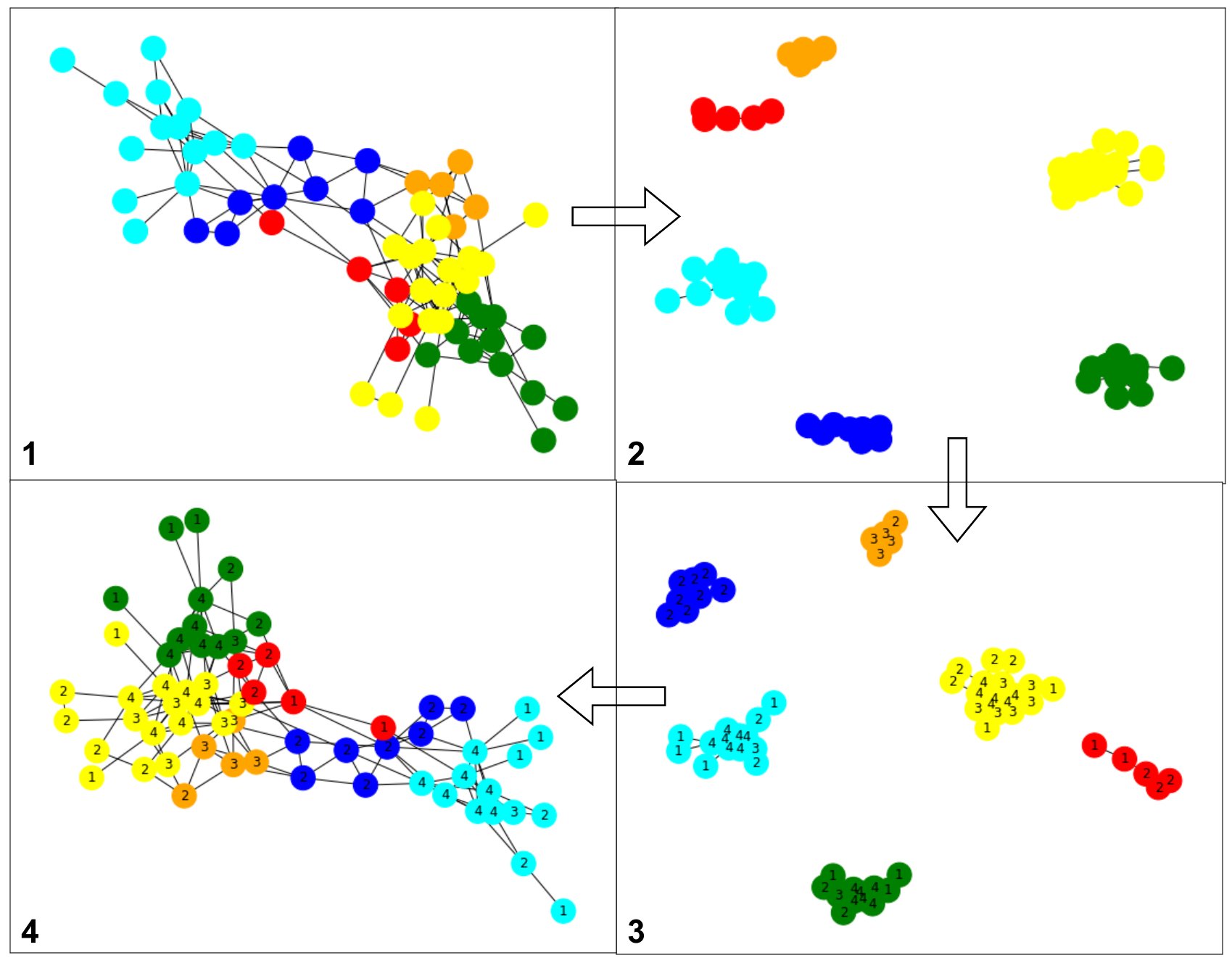}}
\caption{Illustration of the various steps involved in the identification of Community K-shells in a network graph. (1) Graph with communities (2) Isolated community graph (3) CKS in isolated graph (4) CKS in original graph}
\label{fig:cks_toy}
\end{figure*}

Fig. \ref{fig:cks_toy} portrays an illustration of the concept of CKS. The algorithm starts from a network divided into communities identified using the Louvain algorithm. Each community is represented using a different color. These communities are further disconnected from each other using the community isolator (as discussed in subsection \ref{section-cks}), resulting in isolated communities. In order to build the Community K-Shells, these isolated communities are passed through the K-Shell algorithm. Each node is labeled with a number, which represents the respective Community K-Shell (CKS) of the node for the respective community.

\subsection{Calculating K-Shell Entropy (KSE)}
In this step, the connections between a node and its connected community are assigned a KSE score by analyzing their distribution among various CKSs of the community. This aids in determining a node's impact and extent of influence on a given community, if the node becomes activated and acts as a spreader. 
Since most of the real-life communities are composed of a large number of nodes, and an activated node’s impact is believed to last up to only a few hops (2 to 3 hops), just being connected to a given community does not guarantee high influence in that network. Therefore, to ensure the proper spread of information and a strong influence, a node is considered to be a seed node if and only if it has a good number of connections, and the connections are distributed over different regions of the community. 

To ensure the proper distribution of connections of seed nodes, we analyze the distribution of these connections across the community shells generated in the previous step and use Shannon's Entropy over this distribution to check its connectivity to different parts of the networks. A high KSE score conveys that the information reaches the maximum number of shells, and verifies that the connections to that community are not concentrated in one part. It also reduces the possibility of overlapping spread of the same information by nodes in the same locality and ensures that the information spreads in an optimized manner, saving important spreading strength.

Moreover, we constantly ensure that the connections to core nodes are given more importance in comparison to other connections by weighing them with their k-value number, as we can't ignore the fact that the core node would be more impactful individually.

Equation \ref{eq:shells} represents the entropy submission over the shells of the respective community weighted by the K-Value of the respective shell, where $\eta_{v,s}$ is the number of connections of node $v$ with shell $s$ of respective community, $shells_{c}$ represents the unique CKS in the community $c$, and $K_{s}$ represents the k-value of the respective shell.\\
\begin{equation}
\label{eq:shells}
    KSE_{v,c} =  -\sum_{s = 1}^{shells_{c}} K_{s}*\frac{\eta_{v,s}}{\eta_{v}}*log(\frac{\eta_{v,s}}{\eta_{v}})
\end{equation}

\begin{equation}
\label{eq:eta}
    \eta_{v} = \sum_{s'=1}^{shells_{c}}\eta_{v,s'}
\end{equation}

\begin{figure}[h!]
\centering
\centerline{\includegraphics[scale=0.45]{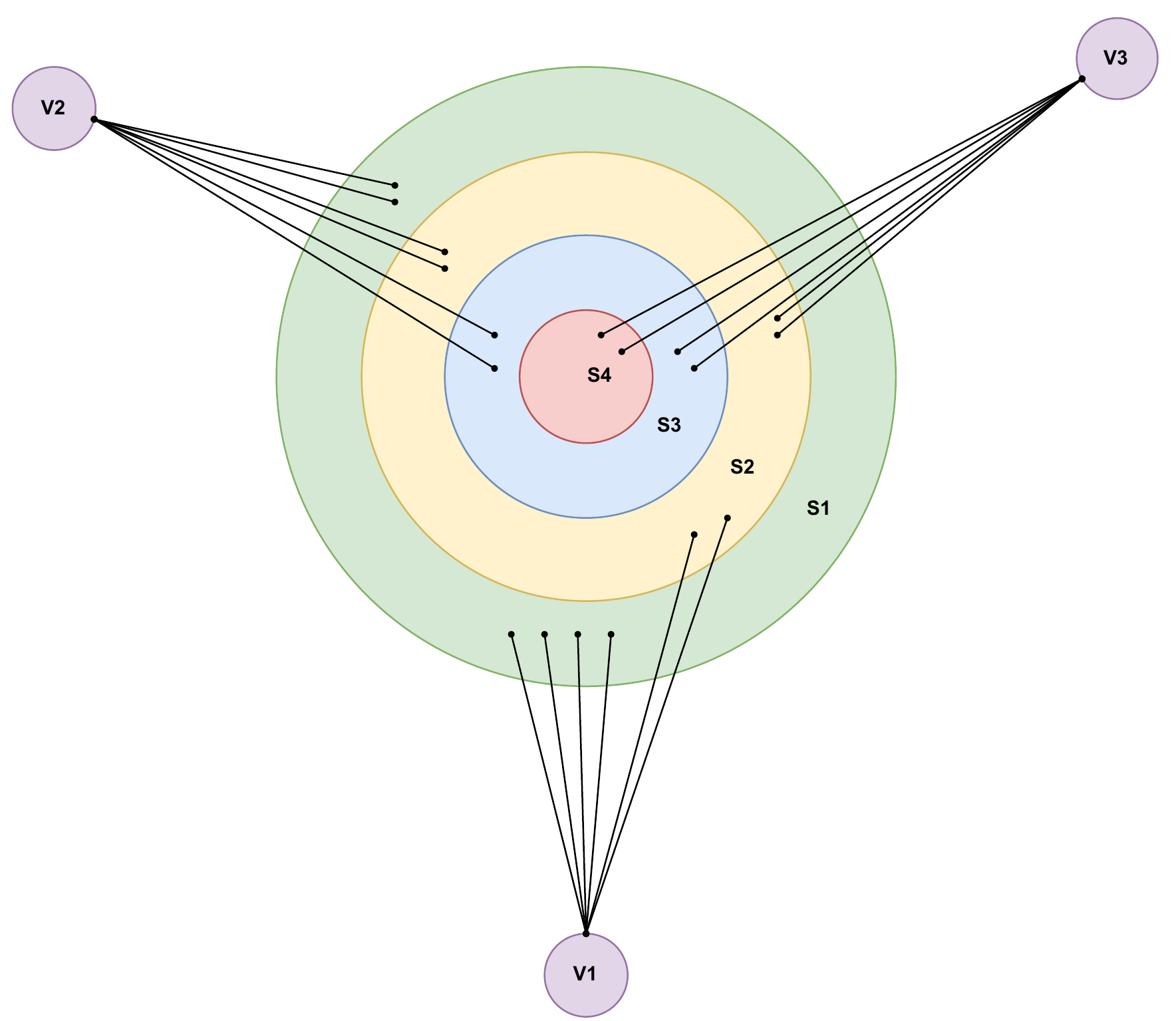}}
\caption{Toy dataset representation of a community with four community k-shells $S_1$, $S_2$, $S_3$ and $S_4$. There exist three nodes $V_1$, $V_2$, and $V_3$, each having six connections with the community, and each distributed differently.}
\label{fig:ckse}
\end{figure}

A better understanding of the concept can be understood from the illustration in Fig. \ref{fig:ckse}. The figure depicts a community that, when passed through the Community K-Shell algorithm, is divided into four shells, namely $S_1$, $S_2$, $S_3$, and $S_4$. We consider three scenarios where each scenario is depicted by each of the nodes $V_1$, $V_2$, and $V_3$, which are connected to the community by six connections distributed among the four community k-shells in different ways. The scoring for each scenario is shown below:\\
\\
\noindent
$V_1 = -1*(4/6)*log(4/6) - 2*(2/6)*log(2/6) = 0.43547500918$\\

\noindent
$V_2 = -1*(2/6)*log(2/6) - 2*(2/6)*log(2/6) - 3*(2/6)*log(2/6) = 0.95424250943$\\

\noindent
$V_3 = -2*(2/6)*log(2/6) - 3*(2/6)*log(2/6) - 4*(2/6)*log(2/6) = 1.43136376416$\\

\noindent
We observe that $V_2$ and $V_3$ scored significantly higher in comparison to $V_1$. This can be attributed to the fact that $V_2$ and $V_3$'s connections are more uniformly distributed as compared to $V_1$'s connections, and since the concept of KSE depends on the concept of entropy, the better the distribution, the higher the score. Further, it was also observed that $V_3$ scored higher than $V_2$, due to the fact that its connections are closer to the core of the community in comparison to $V_2$'s connections, and since the entropy score in KSE is constantly weighed by the shell's k-value, hence the closer the connections to the core, the higher the score. 

\subsection{Calculating CKS score}

In this step, the final influence of a given node is calculated by aggregating the KSE scores of its connections with all the neighboring communities. We consider all the neighborhood communities simultaneously, which gives an overall influence of the node on the whole network. Thus, it is more likely for the algorithm to select a node that spreads information in different communities rather than a single community, hereby ensuring that the selected seed node is connected to more than one community in the network, making it a bridge node.

The scoring also considers the individual characteristics of communities, since each community might have a different impact on the whole network. Hence, the CKS\_score is weighted by community size. Finally, the score for each community is multiplied by the number of connections between the given community and the given node to ensure the absoluteness of the score, since entropy is a relative term. The mathematical equation for the calculation of CKS\_Score is given in Equation \ref{eq:comm}:

\begin{equation}
    \label{eq:comm}
    CKS\_Score(v) = \sum_{c = 1}^{comm} NN_{c}*KSE_{v,c}*\eta_{v}
\end{equation}
Here, $\eta_{v}$ is the number of connections of node $v$ with the respective community as shown in Equation \ref{eq:eta}, $KSE_{v,c}$ represents the K-Shell Entropy for node $v$ and community $c$, and $NN_{c}$ represents the number of nodes in the community $c$.

\begin{algorithm}[h!]
 \caption{: CKS\_SCORE}
 \begin{algorithmic}[1]
 \label{algo:2}
 \renewcommand{\algorithmicrequire}{\textbf{Input:}}
 \renewcommand{\algorithmicensure}{\textbf{Output:}}
 \Require $G = (V,E)$, where $n=|V|$ , $e = |E|$
 
 Communities \ partitioned \ using \ Louvain \ Algorithm
  
$CKS: $Calculated\ using\ Algorithm\ 1 
  
$v: $ Node\ whose\ score\ needs\ to\ be\ computed\ 
  
$K_{s}: $ k-value\ of\ CKS\ in\ a\ community
  
$NN_{c}: $ Number\ of\ nodes\ in\ communities  \setcounter{ALG@line}{0}
  \Ensure  $CKS\_Score$
\State $CKS\_Score \leftarrow 0$
  \For {$c$ $\in$ $\{$communities\ connected\ to\ $v$$\}$}
    \State $KSE_{v,c} \leftarrow -\sum_{s = 1}^{shells_{c}} K_{s}*\frac{\eta_{v,s}}{\eta_{v}}*log(\frac{\eta_{v,s}}{\eta_{v}})$ $(using\ Equations\ \ref{eq:shells}\ \&\ \ref{eq:eta})$
    \State $CKS\_Score \leftarrow NN_{c} * KSE_{v,c} * \eta_{v,c}$  $(using\ Equation\ \ref{eq:comm})$
  \EndFor\\
 \Return $CKS\_Score$
 \end{algorithmic} 
\end{algorithm}

Algorithm 2 describes the computation of CKS\_Score. 
\begin{itemize}
    \item It takes as input the graph $G$, communities ${c_1,c_2,c_3...c_m}$ computed using the Louvain algorithm, CKS scores computed using Algorithm 1, the dictionary of nodes in the community $c$, i.e., $NN_{c}$, and k-value of each CKS shell.
    \item It begins by the initializing score and iterating $c$ over communities connected with $v$. 
    \item Then, it loops over each unique value of CKS present in the community $c$. 
    \item Further, the value of K-Shell Entropy (KSE) for a node $v$ and a community $c$ is calculated by applying the concept of entropy over the ratio of the number of connections between a node $v$ and shell $s$, over the number of connections between node $v$ and all the shells of community $c$. This is further weighted with k-value of each shell, i.e., $K_{s}$, as shown in Equation \ref{eq:shells}.
    \item The K-Shell Entropy of each community is further weighted with the number of connections between node $v$ and the given community $c$, i.e., $\eta_{v}$ and the number of nodes in the respective community $NN_{c}$.\\
\end{itemize}

\section{Datasets, Baseline Models and Performance Metrics} 

In this section, we provide the experimental details of our in-depth analysis done for CKS and competing approaches. We elaborate on the datasets used for evaluation, the baseline approaches for comparison, the performance metrics used, and the experimental setup and parameters.

\subsection{Datasets}

\begin{table}[h!]
    \centering
    \caption{The different datasets used for experimentation along with the various characteristics such as Number of Nodes, Edges and Communities by Louvain Algorithm.}
    \begin{tabular}{|c|c|c|c|}
    \hline
        Dataset & Nodes (n) & Edges (m) & Communities \\ \hline
        Wiki-Vote \citep{wikivote1,wikivote2} & 889 & 2914 & 40\\ \hline
        Twitch \citep{twitch} & 7,126 & 35,324 & 20 \\ \hline
        BA \citep{ba} & 2000 & 9974 & 14 \\ \hline
        Soc Hamsterster \citep{hamsterster} & 2,400 & 16,600 & 169 \\ \hline
        PGP \citep{PGP} & 10,638 & 24,301 & 104 \\ \hline
        PCG \citep{pcg} & 2000 & 9963 & 21 \\ \hline
        p2p-Gnutella04 \citep{gnutella} & 10,876 & 39,994 & 24 \\ \hline
        Email-univ \citep{univ} & 1100 & 5500 & 11 \\ \hline
    \end{tabular}
\label{table-1}
\end{table}

We evaluate our proposed approach on six real-life and two synthetic datasets of varying sizes and types. Table \ref{table-1} describes the different datasets used during our evaluation.

A brief description of different datasets is given below:

\begin{itemize}
\item Wiki-Vote \citep{wikivote1,wikivote2}: The dataset contains Wikipedia voting data right from its inception till January 2008. Nodes represent Wikipedia users and the directed edges from node \textit{u} to node \textit{v} represent that user \textit{u} voted for user \textit{v}.
\item Twitch \citep{twitch}: This dataset is used for node classification and transfer learning based on Twitch user-user networks of gamers. Here, the nodes represent the users themselves, while their mutual friendship is represented by the edges.
\item BA \citep{ba}: BA is a random graph generated using Barabási-Albert preferential attachment model. A network of \textit{n} nodes is generated by adding new nodes, each having \textit{m} edges that are preferentially coupled to existing nodes with a high degree. 
\item Hamsterster \citep{hamsterster}: This network represents the friendships and family links between users on the website hamsterster.com. Here, each user is a node and their friendships are the edges.
\item PGP \citep{PGP}:  Pretty Good Privacy (PGP) is an encrypted communication network. PGP algorithm is used for secure information interchange. Here, the users are nodes, and the edges of this huge network are the connectivity between users.
\item PCG \citep{pcg}: The Power-law Cluster (PCG) \citep{pcg} approach was developed to generate random graphs with a power-law degree distribution and approximate average clustering. The technique requires three parameters (\textit{n}, \textit{m}, \textit{p}), where \textit{n} denotes the number of nodes to be added, \textit{m} denotes the number of random edges to be added for each new node, and \textit{p} is the probability of creating a triangle after adding a random edge.
\item p2p-Gnutella04 \citep{gnutella}: This dataset is one of the snapshots of the Gnutella peer-to-peer file sharing network from August 2002. Nodes represent hosts in the Gnutella network, while edges represent connections between the Gnutella hosts.
\item Email-univ \citep{univ}: This network was generated from a European research institution using their email data from October 2003 to May 2005 (18 months). Here, each node represents an email address, while directed edges represent the flow of emails.
\end{itemize}

\subsection{Baseline Approaches}

The baseline approaches used by us for comparison in this paper are presented below. They have been explained in detail in Section 2.

\begin{itemize}
    \item Extended Neighborhood Coreness (ENC) \citep{ENC}: ENC extends the K-shell decomposition to a two-hop measure by taking the sum of both one and two-hop measures. 
    \item  Gateway Local Rank (GLR) \citep{GLR}: GLR extends over the closeness centrality measure and simplifies it by reducing the search set to the local and gateway nodes.
    \item Degree and Clustering Coefficient and Location (DCL) \citep{DCL}: DCL measures the spreading ability of nodes on the basis of a node's degree, its neighbor, and the clustering coefficient. 
    \item Local Information Dimensionality (LID) \citep{LID}: LID measures the spreading ability of a node by considering the quasilocal structure of a node.
    \item Degree and Importance of Lines (DIL) \citep{DIL}: DIL ranks the nodes on the basis of the degree and the importance of lines.
    \item Betweenness Centrality (BC) \citep{BC}: BC ranks the nodes on the basis of their presence in the shortest path between the nodes. Despite being among the original IM approaches, it still performs comparably with newer approaches.
    \item Closeness Centrality (CC) \citep{CC}: CC ranks the nodes on the basis of their closeness to the rest nodes in the graph. 
\end{itemize}  

\subsection{Performance Metrics}
\begin{itemize}
    \item Final infected scale: The final infected scale is the ratio of infected or active nodes at the end of the spread time and the total number of nodes in the graph. The activation of nodes is caused due to activation probability working in favor of the nodes. The initial number of influential nodes plays a major role in the final number of infected nodes in the network. The fraction of total nodes that act as initial spreaders is called the spreader fraction. Usually, a high value of the spreader fraction leads to a high number of the final infected nodes.
    
    \item Average Shortest Path Length: The Average Shortest Path Length (ASPL) is the average of shortest paths between all the influential nodes when selected in pairs. It helps us determine the efficiency of data transmission in a social network. A higher value of ASPL conveys that the influential nodes are evenly spread and cover a larger part of the network, rather than being concentrated in an area and resulting in a lesser spread in the whole network.
    \item Final infected scale vs Activation Probability (P): Activation probability plays a big role in the total final spread. As the activation probability increases, the final infected scale should also rise. The final infected scale represents the total number of infected nodes at the end of spread time. It plays a major role in the analysis of the effect of activation probability on the total spread of information.
    \item Execution Time: In order to compare our proposed approach with existing techniques, the time taken to rank the nodes on the basis of their spreading ability was also taken into account. The time consumption for all these centralities was evaluated in a Jupyter notebook environment, with the code being run on a Tesla T4 GPU, with 12GB RAM available to the environment.
\end{itemize}

\subsection{Experimental Setup and Parameters}

We evaluated the results of our experiments 100 times for each model on each dataset for each parameter. The results plotted in Fig. \ref{fig:4}, Fig. \ref{fig:5}, Fig. \ref{fig:6}, and Fig. \ref{fig:7} are the average results for each model for 100 simulations of each experiment. 

For calculating the Final Infected State vs Spreader Fraction values for each model, initial spreader fraction values for datasets having less than 2000 nodes were chosen from the set {0.02, 0.03, 0.04, 0.05, 0.06, 0.07, 0.08, 0.09, 0.1}, whereas for larger datasets with greater
than 2000 nodes, the initial spreaders fraction were taken from the set {0.005, 0.01, 0.015, 0.02, 0.025, 0.03, 0.035, 0.04}. For smaller datasets, we have taken relatively larger spreader fractions because smaller values will lead to a very small number of initial spreaders in small datasets. The infection probability, i.e., the probability for a susceptible node to be influenced was taken to be 0.1 for all methods and datasets.

For calculating the Final Infected State vs Activation Probability (P), We took the different activation probability values in the range \{0.05, 0.075, 0.1, 0.125, 0.15, 0.175, 0.2, 0.225, 0.25\}. The initial spreader fraction for each case was taken to be 0.03, and for each dataset, the simulation was performed 100 times. 

While calculating the average spreader distance between initial spreaders, initial spreader fraction values for datasets having less than 2000 nodes were chosen from the set {0.02, 0.03, 0.04, 0.05, 0.06, 0.07, 0.08, 0.09, 0.1}, whereas for larger datasets with greater than 2000 nodes, the initial spreaders fraction were taken from the set {0.005, 0.01, 0.015, 0.02, 0.025, 0.03, 0.035, 0.04}. We set the activation probability (P) to 0.1.

\section{Experimental results and analysis}\label{sec5}

In this section, we present the results of our experiments, as well as their analysis to understand why CKS outperforms competing approaches across all performance metrics while remaining reasonably efficient.

\subsection{Final infected scale}

\begin{figure}[t!]
\centering
  \includegraphics[scale=0.47]{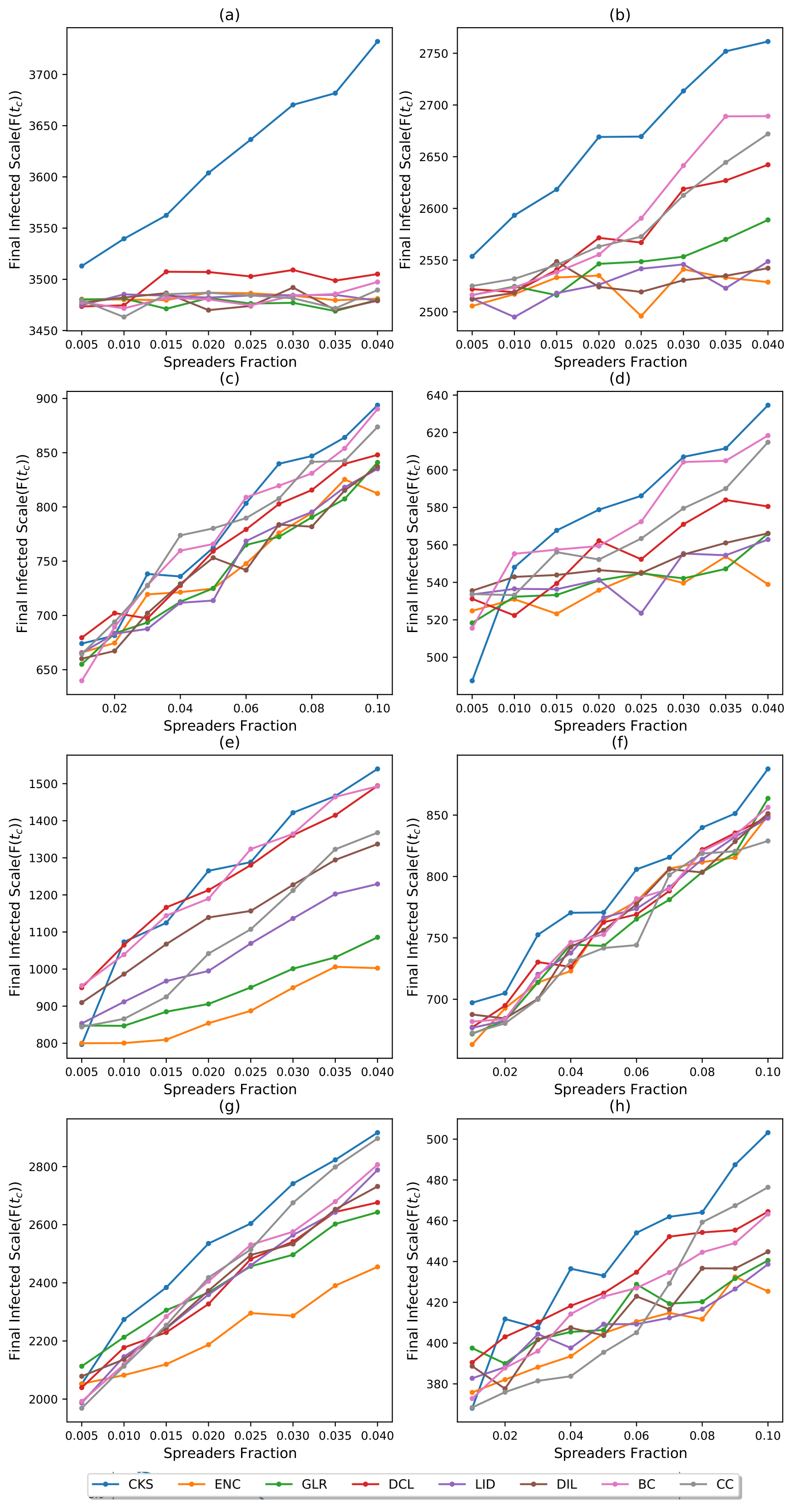}
\caption{Final infected scale plots for various values for initial spreader fraction on the (a) Wiki-Vote dataset (b) Twitch dataset (c) BA dataset (d) Soc-Hamsterster dataset (e) PGP dataset (f) PCG dataset (g) p2p-Gnutella04 dataset (h) Email-univ dataset. The results are averaged for 100 independent simulations of the IC model with an activation probability (P) equal to 0.1.}
\label{fig:4}   
\end{figure}

As discussed in Section 5.3, the final infected scale plots the final infection at the end of the simulation, with respect to the different values for the fraction of initial spreaders. Fig. \ref{fig:4} shows the plots for Final Infected Scale vs Ratio for different approaches. The initial spreader fractions are plotted along the x-axis, while the corresponding final infected values are plotted along the y-axis. We observed that for most of the datasets, final infected numbers increased by increasing the spreader fraction. However, for a few datasets like Wiki-Vote and Twitch, the final infected values did not increase much on increasing the initial spreaders fraction for certain approaches other than CKS. Our proposed approach, CKS, showed significantly better performance as compared to the other novel approaches on all the datasets. This is due to the fact that the seed nodes selected using CKS are farther apart and have a wider influence spread in the network. Other approaches tend to find seed nodes in the same communities, due to which they do not show meaningful improvement in the final infected state upon increasing the spreader fraction. This is due to the overlap of information spread in the network, which occurs lesser for CKS, as its final infected state keeps increasing on increasing spreader fraction values. BC, CC, and DCL also performed reasonably well on all datasets. We also observed that ENC performed the worst among all approaches on this metric. Hence, we can conclude that CKS outperforms other competing methods for a wide range of values for the initial spreader fraction.

\begin{figure}[t!]
\centering
  \includegraphics[scale=0.47]{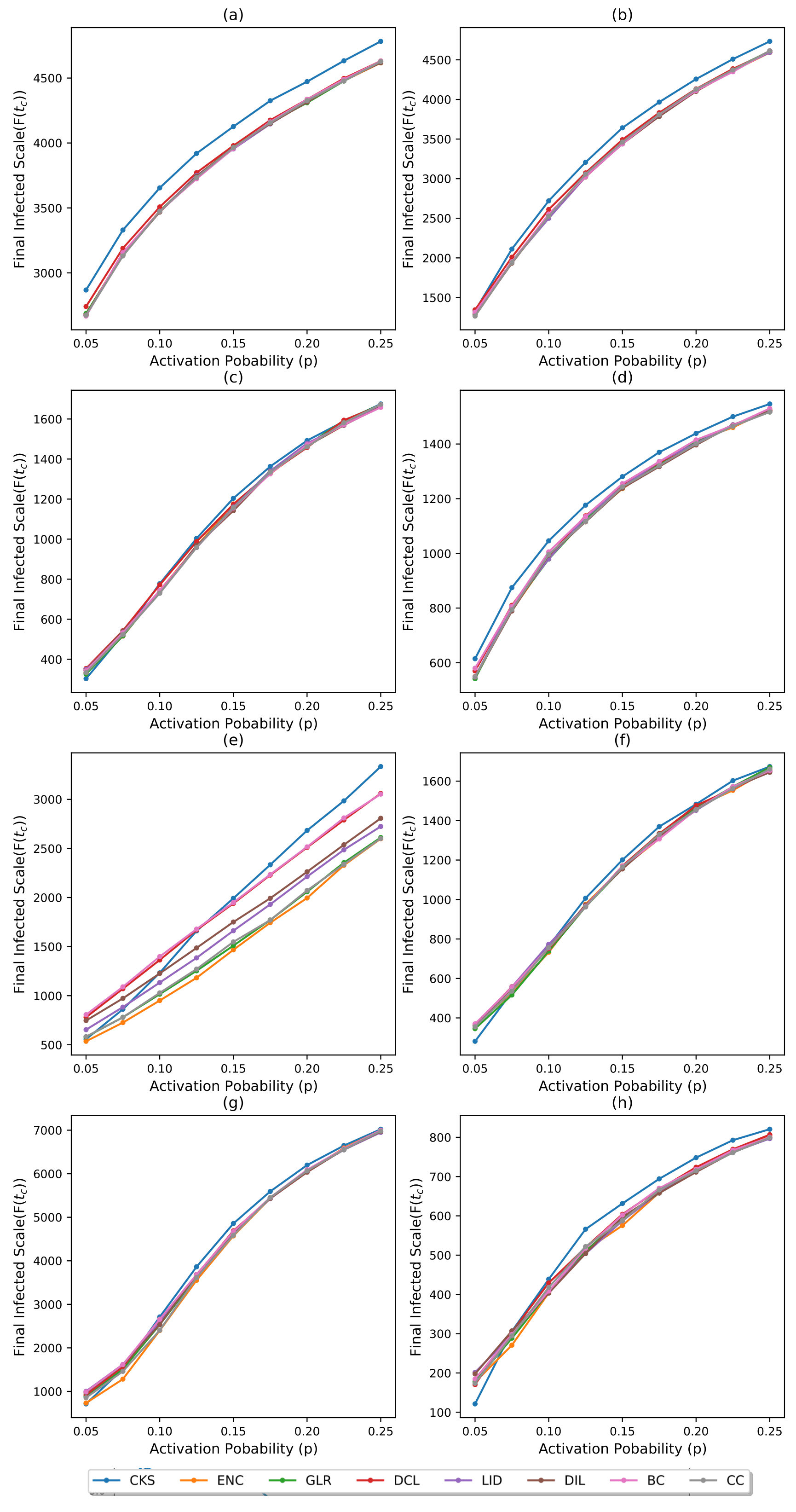}
\caption{Final Infected Scale plots for different activation probability (P) values on (a) Wiki-Vote dataset (b) Twitch dataset (c) BA dataset (d) Soc-Hamsterster dataset (e) PGP dataset (f) PCG dataset (g) p2p-Gnutella04 dataset (h) Email-univ dataset. The results are averaged for 100 independent simulations of the IC model with an initial spreader fraction equal to 0.3.}
\label{fig:5}   
\end{figure}

\begin{figure}[t!]
\centering
  \includegraphics[scale=0.47]{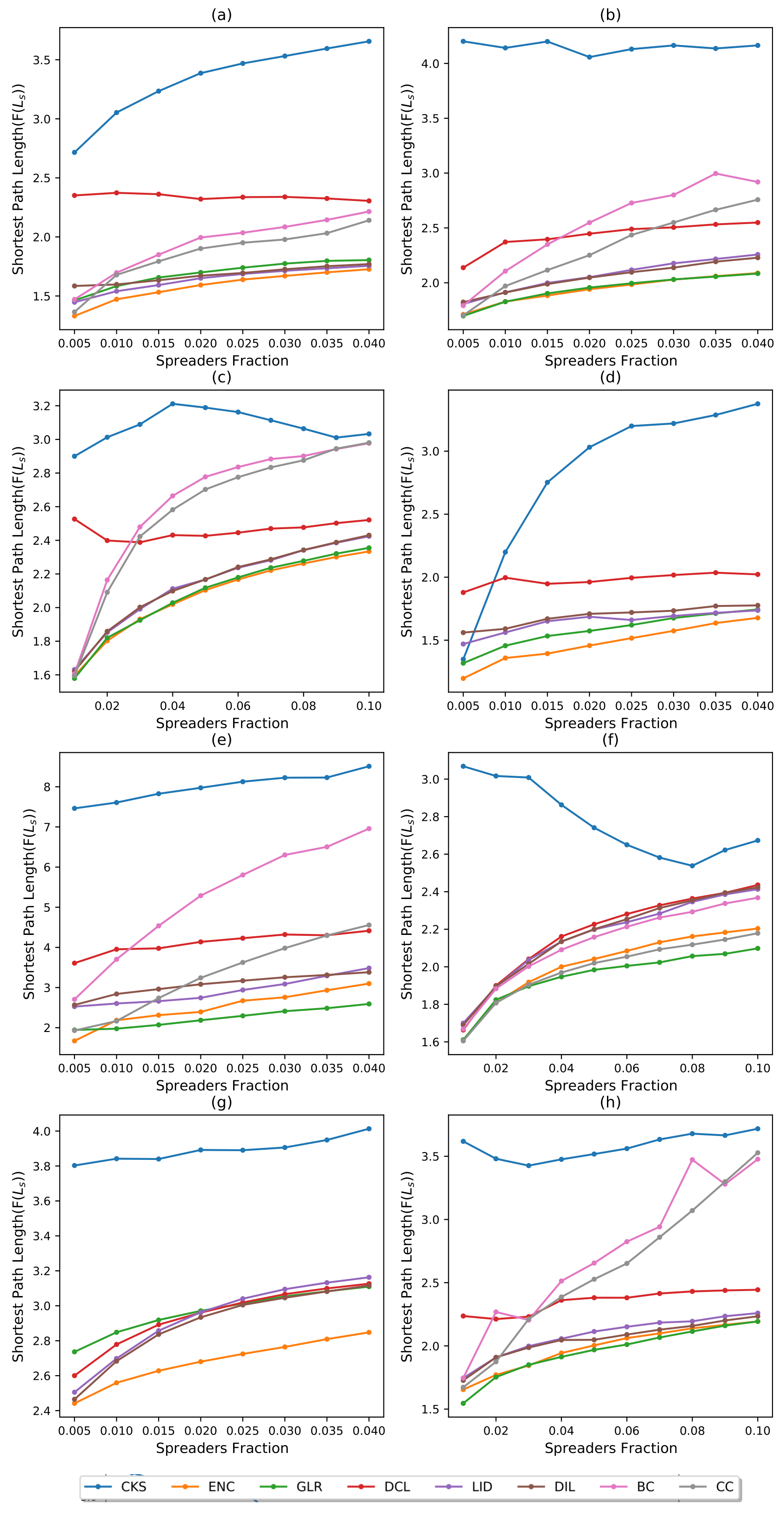}
\caption{Average distance between spreaders with respect to different initial spreaders fraction values for (a) Wiki-Vote dataset (b) Twitch dataset (c) BA dataset (d) Soc-Hamsterster dataset (e) PGP dataset (f) PCG dataset (g) p2p-Gnutella04 dataset (h) Email-univ dataset. The results are averaged for 100 independent simulations of the IC model with an activation probability (P) equal to 0.1.}
\label{fig:6}         
\end{figure}

\begin{figure}[t!]
\centering
  \includegraphics[scale=0.47]{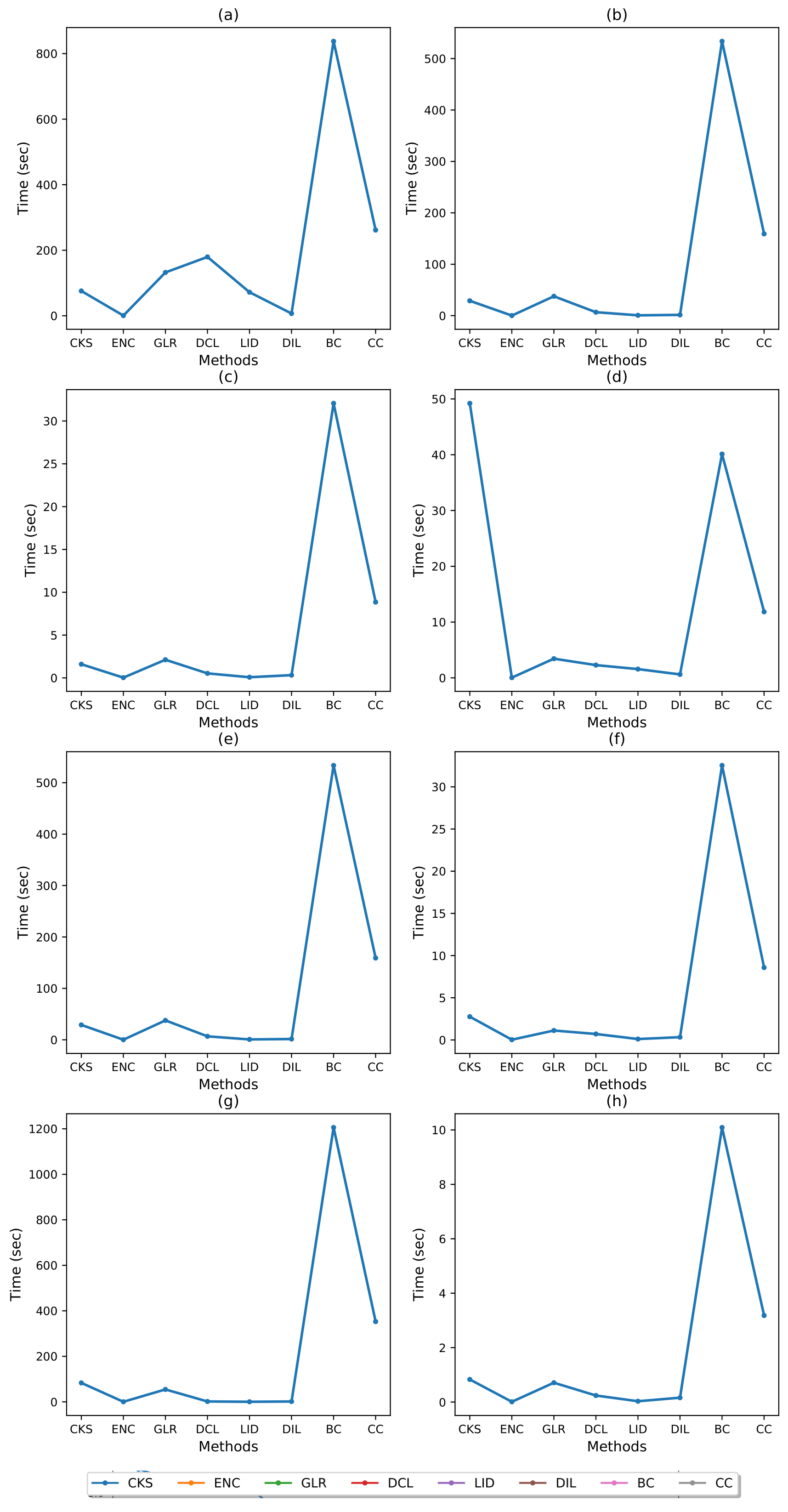}
\caption{The execution time plots for different methods for ranking the spreaders for (a) Wiki-Vote dataset (b) Twitch dataset (c) BA dataset (d) Soc-Hamsterster dataset (e) PGP dataset (f) PCG dataset (g) p2p-Gnutella04 dataset (h) Email-univ dataset.}
\label{fig:7}       
\end{figure}

\subsection{Final infected scale vs Activation Probability (P)}
In this section, we evaluated the performance of our proposed approach for different values of infection probability. Fig. \ref{fig:5} shows the plots for the Final Infected Scale vs Activation Probability for the different methods on all eight datasets. The different values for the activation probabilities are taken along the x-axis, whereas the number of infected nodes is taken along the y-axis. From the given plots, we can observe that as per our expectation, the number of final infected nodes increased upon increasing the activation probability for all methods and across all datasets. CKS consistently performed as well or better than the competing approaches across all datasets, validating its superiority over other approaches. Thus, CKS can propagate information better than other approaches, and this lead only increases on increasing the activation probability.

\subsection{Average Distance between Spreaders}
When selecting the initial spreaders, it is desirable that the selected spreaders should have a minimal overlapping region of influence, such that they are able to influence different parts of the network. To check this, we computed the average shortest distance between the initial spreaders ($L_s$) for different values of spreaders fraction. Fig. \ref{fig:6} illustrates the $L_s$ vs Initial spreaders plots for all eight datasets for our proposed approach as well as other competitive approaches. The x-axis shows the different initial spreader fractions, while the y-axis shows the corresponding value of $L_s$ for each fraction. From the plots for different datasets, we observed that ENC and GLR were among the worst-performing approaches, whereas LID and DIL showed similar performance and were slightly better than ENC and GLR. We also observed that our proposed approach CKS outperformed all other approaches, with DCL and PR being the next best approaches. This indicates a lesser \emph{rich-club effect} in the case of CKS and leads to a wider spread of information in the network. We also observe that CKS as well as other approaches typically reach the maximum average spreader distance around the spreader fraction value of 0.4. Till this point, increasing the spreader fraction only covers other communities which weren't influenced at lower spreader fractions. Beyond 0.4, the density of spreaders in the network increases, leading to a reduction in the average distance between spreaders. However, in almost every situation, CKS outperforms all competing approaches on all datasets, indicating the superiority of CKS.

\subsection{Execution Time}
In this section, we evaluated our proposed approach CKS and other competing approaches on the basis of the execution time required for ranking the nodes for all eight datasets. Their corresponding plots are shown in Fig. \ref{fig:7}. The x-axis conveys the various algorithms, whereas the y-axis shows the execution time in seconds. We observed that BC turned out to be the most computationally expensive method, followed by CC. GLR and CKS were computationally less expensive than CC and BC and took a similar time to execute. ENC took the least amount of time for execution among all the methods, followed by DIL, LID, and DCL. From these plots, we can conclude that our proposed methodology CKS is more efficient than other global centrality measures CC and BC while performing similarly to community-based approaches like GLR. We also found CKS to be computationally more expensive than local hybrid centrality measures like DIL, DCL, and LID.

\section{Statistical Testing}
We further evaluate the performance of our approach CKS using the Friedman statistical test \citep{Friedman}, and Holm's P-values generated from the Iman-Davenport statistic. The former is a non-parametric statistical test that enables multiple comparisons (for two or more methods). It helps us determine if the performance of CKS is noticeably different from other approaches in sets of two or more methods. The Friedman test detects these differences based on the ranking of the methods instead of their errors. It consists of two hypotheses: the null ($H_0$) and the alternate hypothesis ($H_1$). In the former, there are no prominent differences between the algorithms (equality of medians condition), while in the latter, the algorithms have significant differences in the medians of their populations, thus negating the null hypothesis. We performed the Friedman test on the Final Infected Scale vs Spreader Fraction metric using the following procedure:

\begin{itemize}
    \item Gather the generated results for each problem pair.
    \item Rank the values in ascending order from 1 (best value) to $n$ (worst value) for each problem $i$ for a particular algorithm $j$.
    \item For algorithm $j$, the average rank for each problem $i$ is calculated using the Equation \ref{RJ}:
    \begin{equation}\label{RJ}
        R_j = \frac{1}{n} \sum_{i=1}^n(r_i^j)
    \end{equation}
    where $r^j$ is the rank ($1<j<k$) and $R_j$ is the average rank.
    \item Now that all algorithms are ranked according to their priority, compute the Friedman Statistic $F_f$ using the following equation. $F_f$ is based on a Chi-square distribution, with $k-1$ degree of freedom. $F_f$ is computed as shown in Equation \ref{FF}. In the following equation, $n$ represents the number of rows and $k$ represents the number of columns ($n = 8$ and $k=6$). 
    \begin{equation}\label{FF}
        F_f = \frac{12n}{k(k+1)} \left[\sum_{j=1}^k R_j^2 - \frac{k(k+1)^2}{4}\right]
    \end{equation}
\end{itemize}

Table \ref{Stat} shows the rank list calculated for each approach using the Friedman test. CKS clearly has a lower average value, and thus a better rank than competing approaches.

\begin{table}[h!]
    \centering
    \caption{Average ranking of algorithm calculated using the Friedman test}
    \begin{tabular}{|c|c|c|}
    \hline
        Sr No. & Algorithm & Average Ranking \\ \hline
        1 & CKS & \textbf{1.828} \\ \hline
        2 & BC & 3.421 \\ \hline
        3 & DCL & 3.500 \\ \hline
        4 & CC & 4.476  \\ \hline
        5 & DIL & 5.085\\ \hline
        6 & LID & 5.515 \\ \hline
        7 & GLR & 5.789 \\ \hline
        8 & ENC & 6.382 \\ \hline
    \end{tabular}
\label{Stat}
\end{table}

The Friedman test produces relatively conservative results, which is undesirable. Thus, we also use \citet{Iman-Davenport} statistic $F_{id}$, which is computed as shown in Equation \ref{Eq:FID}. It follows the $F-distribution$ with the degree of freedom as $k-1$ and $(n-1)(k-1)$.
    \begin{equation}\label{Eq:FID}
        F_{id} = \frac{(n-1)\chi_F^2}{n(k-1)-\chi_F^2}
    \end{equation}

The unadjusted P-value or Holm P-value, obtained from the Iman-Davenport statistic for the performance of CKS, advocates the rejection of the null hypothesis $H_0$. The computed P-values are below the standard significance level of $\alpha = 0.05$, indicating that there is a prominent difference between the performance of CKS and the baseline approaches. Thus, the obtained P-values help us conclude the negation of the null hypothesis $H_0$. However, they are not appropriate for comparison with different methods. To compare these approaches with each other, we calculated their adjusted P-values with CKS as the control algorithm.

Adjusted P-values (APVs) enable us to draw the correct correlation between these algorithms by taking into account the accumulated family error with respect to the CKS control algorithm. APVs can be directly compared with the significance level $\alpha = 0.5$. In order to calculate the adjusted P-values, we defined a few post-hoc procedures. There exist various post-hoc procedures, such as those proposed by \citet{bonn} and \citet{holland}. They differ from each other in their adjustment of the value of $\alpha$ to compensate for multiple comparisons for multiple methods. In this paper, we have used the common Holm's procedure \citep{Holm} to evaluate the respective APVs. The values of Holm P-values and APVs are always sorted in ascending order. The equation for the same is given below, where indices $i$ and $j$ refers to the main hypothesis whose APVs are being computed and different hypothesis in the set respectively. $P_j$ is the P-value for the $j^{th}$ hypothesis. Holm APV is computed as shown in Equation \ref{Eq:Holm}.

\begin{equation}\label{Eq:Holm}
    Holm APV_i = \min\{v,1\}, \text{where} \,v = \max\{(k-j)p_j:1\leq j\leq i\}
\end{equation}

\begin{table}[h!]
    \centering
    \caption{Adjusted P-values (APVs) using Holm procedure}
    \begin{tabular}{|c|c|c|c|c|}
    \hline
        Sr No. & Algorithm (CKS is control algorithm) & Z score & Holm p-value & Adjusted p-value \\ \hline
        1 & ENC & -10.51 & 3.55e-26 & 2.48e-25 \\ \hline
        2 & GLR & -9.14 & 2.92e-20 &  1.75e-19\\ \hline
        3 & LID & -8.51 & 8.26e-18 & 4.13e-17\\ \hline
        4 & DIL & -7.52 & 2.66e-14  & 1.07e-13\\ \hline
        5 & CC & -6.11 & 4.79e-10 & 1.44e-09 \\ \hline
        6 & DCL & -3.86 & 5.65e-05 & 1.18e-04\\ \hline
        7 & BC & -3.68 & 1.16e-04 & 1.16e-04\\ \hline
    \end{tabular}
\label{Holm}
\end{table}

The results of the APVs using Holm's procedure are shown in Table \ref{Holm}. The intended APVs are less than the significance level, thus rejecting the null hypothesis. Thus, this helps us conclude the superiority of CKS over competing approaches based on two statistical tests as well.

\section{Conclusion}

Social networks have gained tremendous popularity among the masses due to their increasing use cases in today's world. Influence Maximisation is one of the important areas of research on social networks where optimal seed nodes are needed to be chosen for spreading maximum influence. In this paper, we proposed a novel approach for Influence Maximisation. Our approach employed the concepts of community structures, K-Shell algorithm, and Entropy to rank the influencing power of nodes in a network. Our approach can be described as a two-step procedure, where we score the connections between the community and the nodes, and the nodes with the maximum strong connections to significant communities were selected as seed nodes. To check the efficacy of our proposed approach, we simulated our approach on the Information Cascade model over eight datasets, and further compared and evaluated our approach with seven other novel approaches over Final Infected Scale vs Spreader fraction, Final Infected Scale vs Activation Probability, Average Shortest Path Length vs Initial Spreaders, and Execution Time Proficiency metrics. Our proposed approach, CKS, significantly outperformed the existing methodologies in maximizing the spread of information in the network by identifying the most suitable influential seed nodes. CKS also achieves a higher average spreader distance for the same value of spreader fraction as compared to other approaches, indicating a lower rich-club effect and more efficient spread of information in the network.

\appendix

\bibliographystyle{cas-model2-names}

\bibliography{main}

\end{document}